\documentclass[12pt,conference,letterpaper]{IEEEtran}
\usepackage{graphicx}
\usepackage{balance}
\usepackage{amssymb}
\usepackage{lscape}
\usepackage{longtable}

\usepackage[cmex10]{amsmath}
\usepackage{color}
\usepackage{fancyhdr}
\usepackage[caption=false,font=footnotesize]{subfig}

\renewcommand{\thispagestyle}[2]{}

\fancypagestyle{plain}{
        \fancyhead{}
        \fancyhead[C]{first page center header}
        \fancyfoot{}
        \fancyfoot[C]{first page center footer}
}
\begin{document}

\title{scaleBF: A High Scalable Membership Filter using 3D Bloom Filter}

\author{\IEEEauthorblockN{Ripon Patgiri, Sanbuzima Nayak, and Samir Kumar Borgohain}
\IEEEauthorblockA{Dept of Computer Science \& Engineering\\
National Institute of Technology Silchar\\
Assam-788010, India\\
Email: ripon@cse.nits.ac.in, sabuzimanayak@gmail.com, samir@nits.ac.in}}
% \and
% \IEEEauthorblockN{Sabuzima Nayak}
% \IEEEauthorblockA{Dept of Computer Science \& Engineering\\
% National Institute of Technology Silchar\\
% Assam-788010, India\\
% Email: sabuzimanayak@gmail.com}
% \and
% \IEEEauthorblockN{Samir Kumar Borgohain}
% \IEEEauthorblockA{Dept of Computer Science \& Engineering\\
% National Institute of Technology Silchar\\
% Assam-788010, India\\
% Email: samir@nits.ac.in}
% }

\maketitle

\begin{abstract}
Bloom Filter is extensively deployed data structure in various applications and research domain since its inception. Bloom Filter is able to reduce the space consumption in an order of magnitude. Thus, Bloom Filter is used to keep information of a very large scale data. There are numerous variants of Bloom Filters available, however, scalability is a serious dilemma of Bloom Filter for years. To solve this dilemma, there are also diverse variants of Bloom Filter. However, the time complexity and space complexity become the key issue again. In this paper, we present a novel Bloom Filter to address the scalability issue without compromising the performance, called scaleBF. scaleBF deploys many 3D Bloom Filter to filter the set of items. In this paper, we theoretically compare the contemporary Bloom Filter for scalability and scaleBF outperforms in terms of time complexity.
\end{abstract}


\IEEEpeerreviewmaketitle

\section{Introduction}
Burton Howard Bloom introduces a data structure on approximate membership query in 1970 \cite{Bloom}, hence, it is named as Bloom Filter. Bloom Filter is an extensively experimented to enhance a system’s performance since its inception. Moreover, Bloom Filter is also applied numerous areas, namely, Big Data, Cloud Computing, Networking, Security \cite{DDoS}, Database, IoT, Bioinformatics, Biometrics, and Distributed system. However, Bloom Filter is inapplicable in hard real-time system, and password management system \cite{Broder} due to accuracy issues. Applications of Bloom Filter take the lion's share in Computer Networking which includes Named Data Networking (NDN), Content-Centric Networking (CCN), Software-defined Network (SDN), Domain Name System (DNS), and Computer Security. The Bloom Filter reduces space consumption in an order of magnitude as compared to a conventional hash algorithm. However, Bloom Filter cannot stand itself. Bloom Filter is used as enhancer of a system. For example, BigTable uses Bloom Filter to reduce the number of disk accesses which improves the performance drastically \cite{Chang}. Similarly, in Cassandra \cite{Lakshman}.

\subsection{Motivation}
Several variants of Bloom Filters have been developed to address some issues \cite{RS}. However, most of the Bloom Filters are developed to address scalability issue. Guanlin Lu et al. \cite{Lu} proposes a Forest-structured Bloom Filter (FBF). The FBF is a combination of RAM and flash memory. Similarly, Debnath et al. \cite{BloomFlash} develops a very high scalable Bloom Filter based on RAM and flash memory. BloomStore is also another highly scalable Bloom Filter \cite{GLu}. However, these solutions are hierarchical, and thus, lookup and insertion cost is very high. It takes $O(logn)$ time complexity in insertion and lookup operations as demonstrated in Table \ref{tab}.
\subsection{Contribution}
\begin{table*}[!ht]
    \centering
    \caption{Comparison of various scalable Bloom Filter}
    \begin{tabular}{p{2cm}p{2cm}p{2cm}p{2cm}p{2cm}p{1.5cm}p{2cm}}
    \\ \hline
    \textbf{Name} & \textbf{Types} & \textbf{Insertion} & \textbf{Lookup} & \textbf{Scalability} & \textbf{Platform}& \textbf{Algorithm}\\
    \hline
    BloomFlash \cite{BloomFlash} & Hierarchical & Logarithmic & Logarithmic & High & RAM \& Flash & Serial\\
    FBF \cite{Lu} & Hierarchical & Logarithmic & Logarithmic & High & RAM \& Flash & Serial\\
    BloomStore \cite{GLu} & Linear Chaining & Constant & Constant & High & RAM \& Flash & Parallel\\
    TB$^2$F \cite{TB2F} & Hierarchical & Logarithmic  & Logarithmic & Medium & RAM & Parallel\\ 
    Bloofi \cite{Bloofi} & Hierarchical &  Logarithmic  & Logarithmic & Medium & RAM & Serial\\
    scaleBF & 3D & Constant & Constant & High & RAM & Serial\\
    \\ \hline
    \end{tabular}
    \label{tab}
\end{table*}

To address scalability issues, we propose a novel scalable Bloom Filter, called scaleBF. scaleBF is a very simple data structure yet powerful. scaleBF increases its scalability without compromising the performance. scaleBF takes $O(1)$ time complexity in lookup and insertion operations, which is compared in Table \ref{tab}. scaleBF uses chaining hash data structure for implementing the scalability. Also, scaleBF deploys 3DBF \cite{rDBF} to inherit the performance and low memory consumption.

Table \ref{tab} depicts the most scalable Bloom Filters. BloomFlash \cite{BloomFlash}, and FBF \cite{Lu} uses hierarchical structures to indexed the Bloom Filters. BloomStore \cite{GLu} uses linear chain data structure (not open hashing data structure) to store the Bloom Filters in Flash memory. Moreover, BloomStore is designed to perform parallel lookup operation. On the contrary, scaleBF uses chaining hash data structures to achieve higher scalability without compromising the performances. TB$^2$F \cite{TB2F} deploys tree-bitmaps and Bloom Filter, and used for name lookup in Content-Centric Network (CCN). The input is split into a T-segment of fixed size and a B-Segment of variable size. The T-segment key is inserted into bitmap-trie, and the B-segment is inserted into Bloom Filter. However, maintaining trie data structure is costly in terms of space as well as time. On the other hand, Bloofi \cite{Bloofi} uses tree structured Bloom Filter which cuases costly in insertion and lookup. The scalability of BloomFlash \cite{BloomFlash}, FBF \cite{Lu}, BloomStore \cite{GLu}, scaleBF is higher than TB$^2$F and Bloofi \cite{Bloofi}.
\subsection{Organization}
The article is organized as follows- Section \ref{PS} presents the proposed system, called scaleBF. The architecture of scaleBF is demonstrated in Section \ref{PS}. Section \ref{Ana} presents a theoretical analysis on scaleBF. Also, every aspect of scaleBF is analyzed in Section \ref{Ana}. Article discusses cons of scaleBF in Section \ref{Dis}. Finally, the article is concluded in Section \ref{Con}. 
\section{{scaleBF}: The Proposed System}
\label{PS}
\subsection{3D Bloom Filter (3DBF)}
The 3-Dimensional Bloom Filter (3DBF) is similar to conventional Bloom Filter except array structure \cite{rDBF}. The 3DBF uses 3D arrays and it is a static Bloom Filter in nature. The static Bloom Filter does not change the size at run time. Also, static Bloom Filter does not readjust with ever growing data. However, a new 3DBF is created to address the scalability issue.  
\begin{figure}[!ht]
    \centering
    \includegraphics[width=0.35\textwidth]{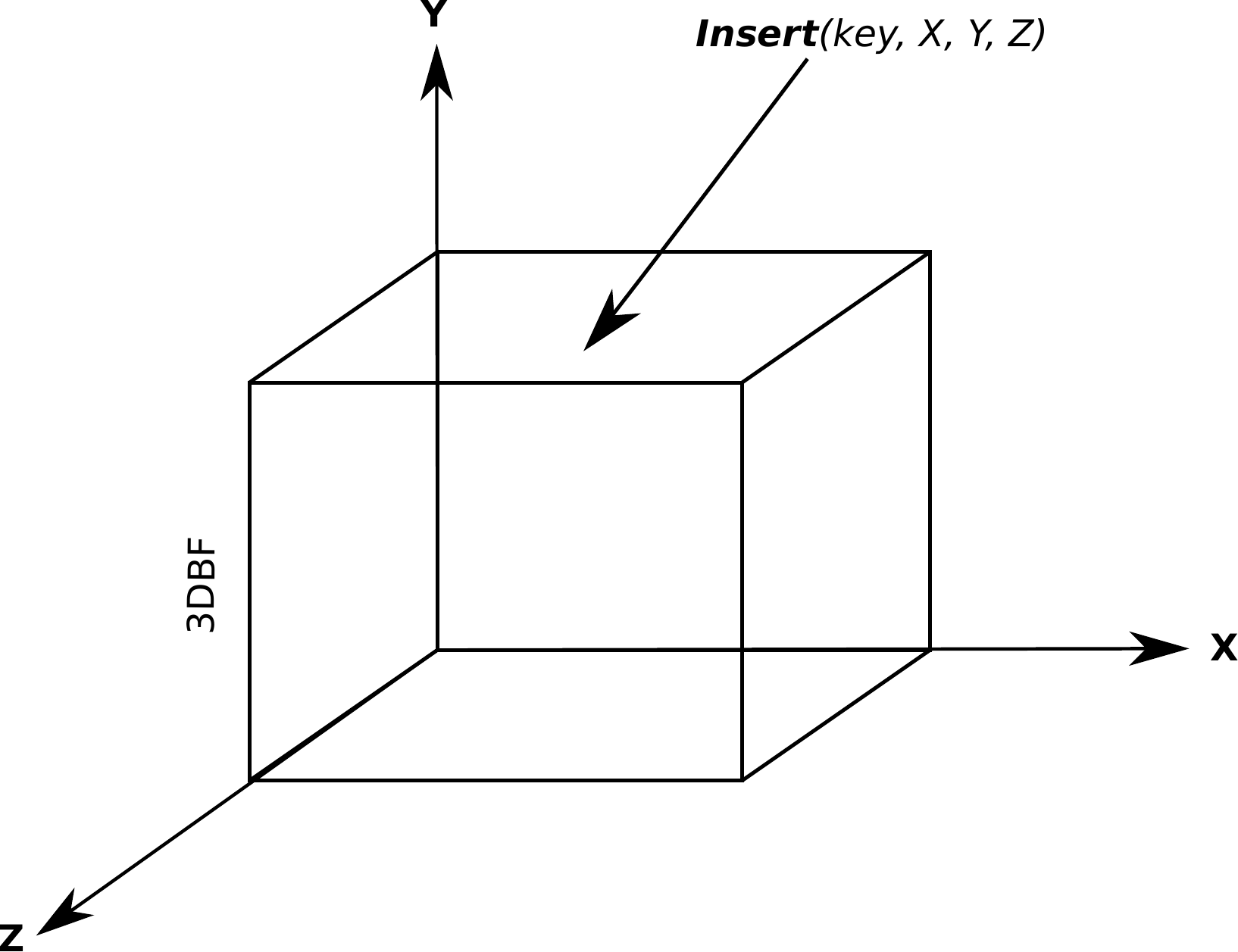}
    \caption{3DBF architecture}
    \label{3d}
\end{figure}

Figure \ref{3d} depicts the architecture of 3DBF. The 3D Bloom Filter uses four modulus operator using prime numbers instead of hashing a key into $k$ different places. These modulus reduces the false positive probability. Thus, 3DBF is independent from number of hash functions $k$. Let, $\mathbb{B}_{X,Y,Z}$ bet the 3DBF where $X,~Y$ and $Z$ be the dimension of the filter. The dimensions are prime numbers, otherwise, false positive increases. Let, $\mathbb{B}_{i,j,k}$ be a cell of the 3DBF. The cell stores \textbf{long int} which occupies $64-bits$. Let us insert a key $\kappa$. The 3DBF uses Murmur hashing \cite{murmur} to generate a hash-value of input item $\kappa$. Let, $h$ be the generated hash-value by Murmur hashing. Now, $i=h\%X,~j=h\%Y,~k=h\%Z$, and $\rho=h\%63$, where $\rho$ is the bit position of the cell $\mathbb{B}_{i,j,k}$. 3DBF sets a bit using Equation \eqref{eq1}-
\begin{equation}\label{eq1}
\mathbb{B}_{i,j,k}\leftarrow \mathbb{B}_{i,j,k}~OR~(1<<\rho)
\end{equation}
where $OR$ is bitwise OR operator. Equation \eqref{eq1} is invoked to insert an input item into 3DBF. The lookup operation requires similar calculation. Equation \eqref{eq2} is invoked to perform the lookup operation in a 3DBF.
\begin{equation}\label{eq2}
Flag\leftarrow (\mathbb{B}_{i,j,k}\oplus (1<<\rho))AND(1<<\rho)
\end{equation}
If $Flag$ is assigned by `1', then 3DBF returns true, otherwise, it returns false. Each item requires a single bit in 3DBF as disclosed in Equation \eqref{eq1}, and each cell has $63-bits$. Therefore, 3DBF consumes the lowest memory as compared to other variants of Bloom Filter. Moreover, 3DBF features detection of the fullness of the filter. 3DBF defines the criticality factor to consider whether the filter is full or not \cite{rDBF}.  

\subsection{Insertion operation in scaleBF}
scaleBF deploys chaining mechanism of conventional hashing data structure for highly scalable. scaleBF deploys many 3DBFs.
\begin{figure*}[!ht]
    \centering
    \includegraphics[width=0.9\textwidth]{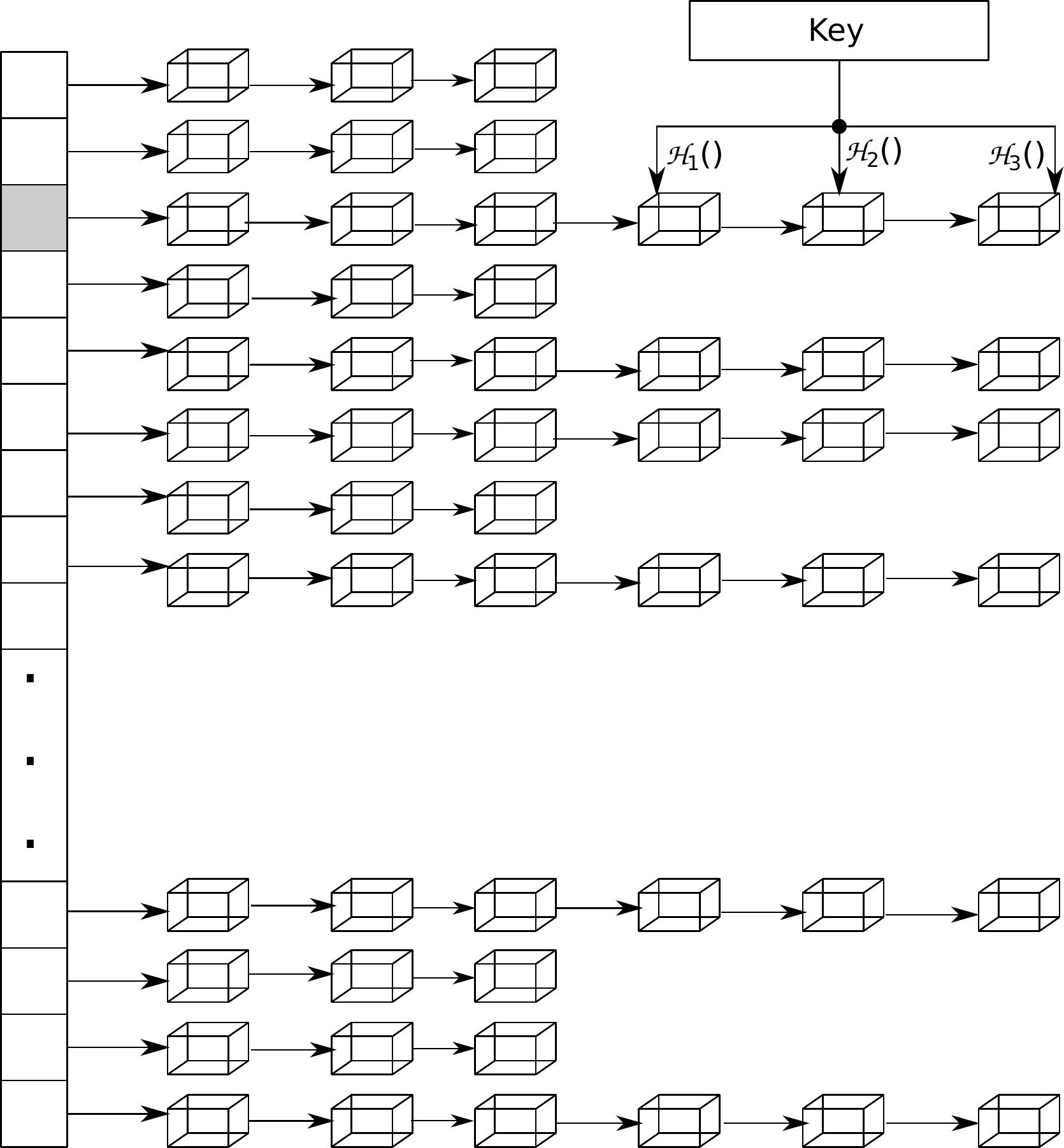}
    \caption{Insertion of an input item and increment of filter size using conventional chaining in scaleBF.}
    \label{ins}
\end{figure*}
\subsubsection{Insertion of Bloom Filter}
A Bloom Filter is formed by three 3DBFs. Each Bloom Filter is formed by three 3DBF. However, Bloom Filter can be formed by augmenting more 3DBF, but we have chosen three for simpler illustration. Each key is inserted into three 3DBF. Let, $\eta$ be the chain size, and input item $\kappa$ to be inserted. There are $\eta$ chains in scaleBF. A new Bloom Filter (three 3DBF) is inserted into the chain if the Bloom Filter (three 3DBF) in particular chain is full. 
\subsubsection{Insertion of a Key}
Insertion of the key is performed using Equation \eqref{eq1} and hashes the key into the particular chain. If a Bloom Filter (three 3DBF) size is full, then move to the last Bloom Filter (three 3DBF). Insert the key using Equation \eqref{eq1}. A key is hashed into particular slot of the chain. There are many Bloom Filters in the particular slot linked with each other as shown in Figure \ref{ins}. If first three Bloom Filter is full, then create and link three 3DBF as demonstrated in the figure. 

\subsection{Lookup operation in scaleBF}

\begin{figure*}[!ht]
    \centering
    \includegraphics[width=0.9\textwidth]{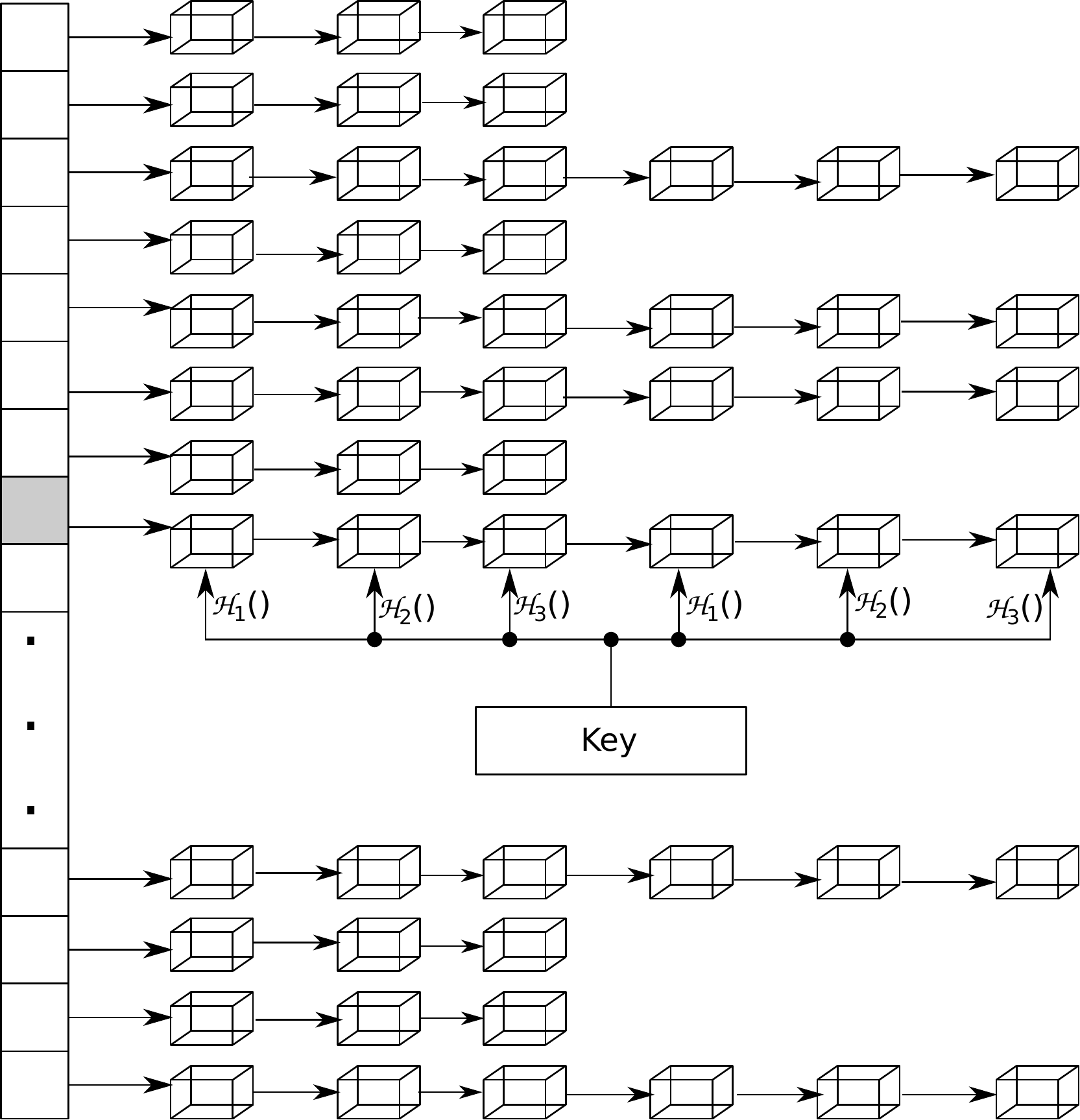}
    \caption{Lookup of an input item in scaleBF.}
    \label{lu}
\end{figure*}
Figure \ref{lu} depicts the lookup operations of the scaleBF. A key is hashed into particular chain and lookup all Bloom Filters sequentially. As a comparison, three 3DBF is searched. If the first three Bloom Filter returns true, then the key is member of Bloom Filter. Otherwise, move forward to the next three 3DBF, and so on. 

\section{Analysis}
\label{Ana}
There is no significant difference between 3DBF and conventional Bloom Filter analysis of number bits consumed, except $k=1$ in 3DBF. Therefore, scaleBF is analyzed through the conventional Bloom Filter. Let, $m$ be the size of Bloom Filter, $n$ be the number of entries, and $k=1$ be the number of hash function, then the probability of a bit to be `0' is
\[\left(1-\frac{1}{m}\right)^n\]
Therefore, probability of total bit to be `1' is
\[\left(1-\left(1-\frac{1}{m}\right)^n\right)\]
Since, scaleBF uses 3DBF, thus, $m=X\times Y\times Z \times 63$.
F. Grandi \cite{Grandi} present a new way to calculate the false positive probability using $\delta-transformation$. Let, $X$ be the random variable to represent the total number of `1' in the Bloom Filter, then
\[E[X]=m\left(1-\left(1-\frac{1}{m}\right)^n\right)\]
The probability of false positive is conditioned to a number by $X=x$, then
\[Pr(FP|X=x)=\left(\frac{x}{m}\right)\]
Therefore, false positive probability is
\[FPP=\sum_{x=0}^m Pr(FP|X=x) Pr(X=x)\]
\[FPP=\sum_{x=0}^m \left(\frac{x}{m}\right) f(x)\]
where $f(x)$ is probability mass function of $X$. F. Grandi \cite{Grandi} applies $\delta-transformation$ to calculate $f(x)$ and presents $FPP$ as follows-
\begin{equation} \label{eq4}
FPP=\sum_{x=0}^m \left(\frac{x}{m}\right) \dbinom{m}{x} \sum_{j=0}^x \left(-1\right)^j \dbinom{x}{j} \left(\frac{x-j}{m}\right)^n
\end{equation}
Equation \ref{eq4} is calculation of false positive probability of a 3DBF. scaleBF deploys three 3DBF. Therefore, the false positive probability of a Bloom Filter having three 3DBF is
\begin{equation}\label{eq5}
    FPP=\prod_{p=1}^3\left(\sum_{x=0}^m \left(\frac{x}{m}\right) \dbinom{m}{x} \sum_{j=0}^x \left(-1\right)^j \dbinom{x}{j} \left(\frac{x-j}{m}\right)^n\right)
\end{equation}
Let, there are $n$ Bloom Filter (three 3DBF each), and their false positive probabilities are $FPP_i$, where $i=1,2,3,\ldots,n$. From Equation \eqref{eq5}, the average false positive probability of scaleBF is 
\begin{equation}\label{eq6}
    FPP_{avg}=\frac{\sum_{i=1}^m~FPP_i}{n}
\end{equation}
Equation \eqref{eq6} presents the false positive probability of scaleBF. 
\subsection{Scalability}
Scalability is the key barrier to the modern Bloom Filter. There are numerous Bloom Filter that addresses the scalability issue. However, scalable Bloom Filters are developed based on reordering, hierarchical and forest structure. scaleBF uses simple hashing scheme to enhance the scalability of Bloom Filter. The chaining is the most used hashing data structure. However, chaining has linear search in the worst case, i.e., $O(n)$ time complexity. In other words, all keys are hashed into single chain location. However, it is once in a blue moon in real-world. Besides, most of the chain remains unused. Therefore, the chaining size must be a prime number to avoid the above situation. 

Undoubtedly, the scalability is achieved using chaining data structure in scaleBF. The RAM size of the system also plays a vital role in scaleBF. 3DBF allocates memory dynamically which requires few memory blocks be contiguous to satisfy the request by the most modern programming language. Therefore, there is less worry about the unavailability of memory blocks. However, scaleBF does not guarantee the availability of the memory. 

Let, $P$ be the slot size and $Q$ be the number of chains to be stored in chaining. The load factor $\alpha=\frac{Q}{P}$, where $P$ is a prime number, and $Q$ is the total Bloom Filter to be inserted. Therefore,
\begin{equation}\label{eq7}
    Q=\sum_{i=1}^T \frac{m_i}{3}
\end{equation}
where $m_i$ is the size of $i^{th}$ 3DBF. Then, the load factor becomes
\begin{equation}\label{eq8}
    \alpha=\frac{\sum_{i=1}^T \frac{m_i}{3}}{P}
\end{equation}
The total available bits in scaleBF are 
\begin{equation}\label{eq9}
    \frac{\tau\times X\times Y\times Z\times\left(\sum_{i=1}^T \frac{m_i}{3}\right)}{P}
\end{equation}
where $\tau$ is the threshold that depends on the requirements, $X,~Y$, and $Z$ are the dimensions. The $\tau$ is calculated by $1,2,3,\ldots,\beta$ and $\beta$ be the number of bits per cell in a 3DBF \cite{rDBF}. For high accuracy, $\tau$ is set to $1$. However, $\tau=\beta$ defines that false positive is insignificance.
\subsection{Time and Space Complexity}
The time complexity is also a key barrier in the scalable Bloom Filter. Hierarchical Bloom Filter or Forest Structured Bloom Filter takes $O(logn)$ time complexity in lookup and insertion operation. Other variants of scalable Bloom Filters also decrease the performance. scaleBF uses $O(1)$ time complexity to lookup and insertion operation on an average case. However, the worst case time complexity is $O(n)$ and it is impractical. 

Let, a key $\kappa$ to be inserted into scaleBF. The $\kappa$ is hashed into a particular slot of chain and insert into the key $\kappa$ in desired Bloom Filter (three 3DBF). If the first Bloom Filter is full, then move to the next and so on. Let, the maximum, the size of a particular chain is $\mathcal{C}$. scaleBF uses prime number $P$ to evenly distribute the keys as disclosed in Equation \eqref{eq8}. Thus, the size of $\mathcal{C}$ is small. Let us, there are $70\%$ slots empty even if prime number $P$. That is, $30\%$ slots are filled. Then, each slot has at least $30\%$ of $Q$ which is also very small. However, the $P$ is a prime number, and thus, the distribution is fair enough to fill each slot. Thus, $\mathcal{C}$ is very small and the total time complexity is $O(1)$ on an average. Similarly, lookup cost also $O(1)$ on an average case.

\subsection{Performance}
scaleBF also inherits the performance of 3DBF \cite{rDBF}. The insertion and lookup cost depends on the cost of Equation \eqref{eq1} and \eqref{eq2}. Equation \eqref{eq1} and \eqref{eq2} uses Murmur hashing \cite{murmur}, which is known as a very fast string hashing. The computational complexity of Murmur hashing is $O(1)$, since, the length of a string is constant and small. Therefore, the Equation \eqref{eq1} and \eqref{eq2} also cost  $O(1)$ time complexity. 3DBF enhances the performance by reducing the total number of complex arithmetic operations. Thus, scaleBF increase its scalability without compromising the performance. 

\section{Discussion}
\label{Dis}
scaleBF provides impressively very high scalability. However, the initial cost of memory consumption can be high. For instance, insert a key which mapped to the slot $3$ of chaining, and creates new three 3DBF. Another insertion key also triggers creation of new three 3DBF which is mapped into a slot, say 2. Thus, the initial cost of memory is high. However, scaleBF is ideal for very large scale membership filtering. Moreover, scaleBF also ideal solution of large memory allocation due to dynamic memory allocation system. scaleBF also depends on the size of 3DBF.

\section{Conclusion}
\label{Con}
Deduplication requires very high scalable Bloom Filter, since, deduplication processes trillions of keys. Moreover, there are diverse applications of high scalable Bloom Filter, for instance, DNA Assembly. In this paper, we have presented a very high scalable Bloom Filter without comprising the performances. In addition, scaleBF also provides insertion and lookup cost of $O(1)$. scaleBF outperforms Bloofi \cite{Bloofi}, BloomFlash \cite{BloomFlash}, FBF \cite{Lu}, and TB2F \cite{TB2F} in terms of computational time complexity while maintaining higher scalability. However, the scaleBF does not support deletion of an item. Thus, there is no false negative. Interestingly, scaleBF can be applied many research areas to boost up the performance and scalability, and its applicability not limited to NDN, but also Big Data, Cloud Computing, Database, Distriubuted System, IoT, and Computer Networking.

\balance
\bibliographystyle{IEEEtran}
\bibliography{mybibfile}

\end{document}